\documentstyle [epsf,epsfig,11pt]{article}  
\def\ra{\rightarrow}

\def\be{\begin{equation}}
\def\ee{\end{equation}}
\def\bea{\begin{eqnarray}}
\def\eea{\end{eqnarray}}

\newcommand{\Mm}{$\mbox{MeV}/\mbox{c}^2$}

\newcommand{\Gm}{$\mbox{GeV}/\mbox{c}^2$}

\newcommand{\Pb}{${\mbox{pb}}^{-1}$}
\newcommand{\Fb}{${\mbox{fb}}^{-1}$}
\newcommand{\Ll}{${\mbox{cm}}^{-2} \ {\mbox{s}}^{-1}$}

\def\pt{$p_T$}
\def\Et{$p_T$}
\def\etae{$|\eta|$}

\def\Pt{$P_T$}
\def\Dzero{D\O\ }
\def\pp{$p\bar{p}$}

\def\ttb{$\mbox{t}\bar{\mbox{t}}$}

\begin{document}

\vspace*{-1.8cm}
\hspace*{13cm}{\bf LAL 99-44}\\
\vspace*{-0.5cm}
\hspace*{13.3cm}{October 1999}
\vskip 3.5 cm

\begin{center}
{\LARGE\bf \Dzero Upgrade for RUN II}
\end{center}

\vskip 1 cm
\begin{center}
{\Large\bf P. Petroff}
\end{center}

\vspace*{0.5cm}

\begin{center}
{\large\bf Laboratoire de l'Acc\'el\'erateur Lin\'eaire},\\
IN2P3-CNRS et Universit\'e de Paris-Sud, BP 34, F-91898 Orsay Cedex\\
{\scriptsize E-mail: pierre.petroff@cern.ch}
\end{center}

\vskip 1 cm
\begin{abstract}
The \Dzero detector at The Fermilab Tevatron is undergoing a major upgrade to prepare for data taking with luminosities reaching $2 \times 10^{32}$ \Ll. The upgrade includes a new central tracking array, new muon detector components and electronic upgrades to many subsystems. The \Dzero upgraded detector will be operational for RUN II in spring 2000.
\end{abstract}
\vskip 1 cm
\section{Introduction}
The Fermilab \pp \ collider remains the highest energy accelerator since mid 80's. He has delivered to CDF and \Dzero detectors an integrated luminosity of 120 \Pb during the Run I period, between 1992 and 1996, at a peak luminosity up to \mbox{$1.6 \times 10^{31}$ \Ll}.
The major studies encompass the top quark discovery, a precise measurement of the W mass, improved limits on tribosons anomalous coupling, improved limits on
SUSY and more exotic particules masses and important tests of QCD in high \Pt  jets production and b quark production.

The next run in collider mode is scheduled to begin in spring 2000 with important improvements to the accelerator complex and to the \Dzero detector. By the end of the Run II, an integrated luminosity of several \Fb is expected. The gain in luminosity by more than an order of magnitude coupled with the upgrade of the detector, will allow more precise measurements in the Standard Model domain and will increase the potential for new phenomena discovery.

\section{The physics program}
The physics program is the continuation of the Run I program with some shifts on the emphasis, due to the higher luminosity and to the better quality of the data. It is  beyond the scope of this  paper to present an
exhaustive    description of  the  physics   program  which will  be  undertaken~\cite{tev2000}. A somewhat  arbitrary  choice of the  "hottest"  subjects are presented:

\begin{itemize}
\item Exploration of the mechanism of the electroweak symmetry breaking with a large sample of top and precise measurement of the EW parameters;
\item Direct searches for new phenomena beyond the Standard Model;
\item CP violation in the b-quark sector;
\item Precise studies of QCD. 
\end{itemize}

\subsection {Properties of the top quark}
 \begin{itemize}
 \item  \textbf{\ttb~production}. With a sample of over 1000 identified 
         b-tagged events: measurement of M$_{top}$ to $\pm$ 3 GeV;
          measurement of $\sigma_{t\bar{t}}$ to $\simeq 10\%$;
       detailed studies of top decays (branching ratios, FCNC, 
              polarization of the W).
 \item \textbf{Single top production}. With a sample of $\simeq 400$ events
     determine 
        $\frac{\delta\sigma(t)}{\sigma(t)}$ to  $\simeq 25\%$ and 
        $\vert V_{tb} \vert$  to $14\%$.
  \end{itemize}

\subsection {Precise measurements of Electroweak parameters}

\begin{figure}
\begin{center}
\epsfig{file=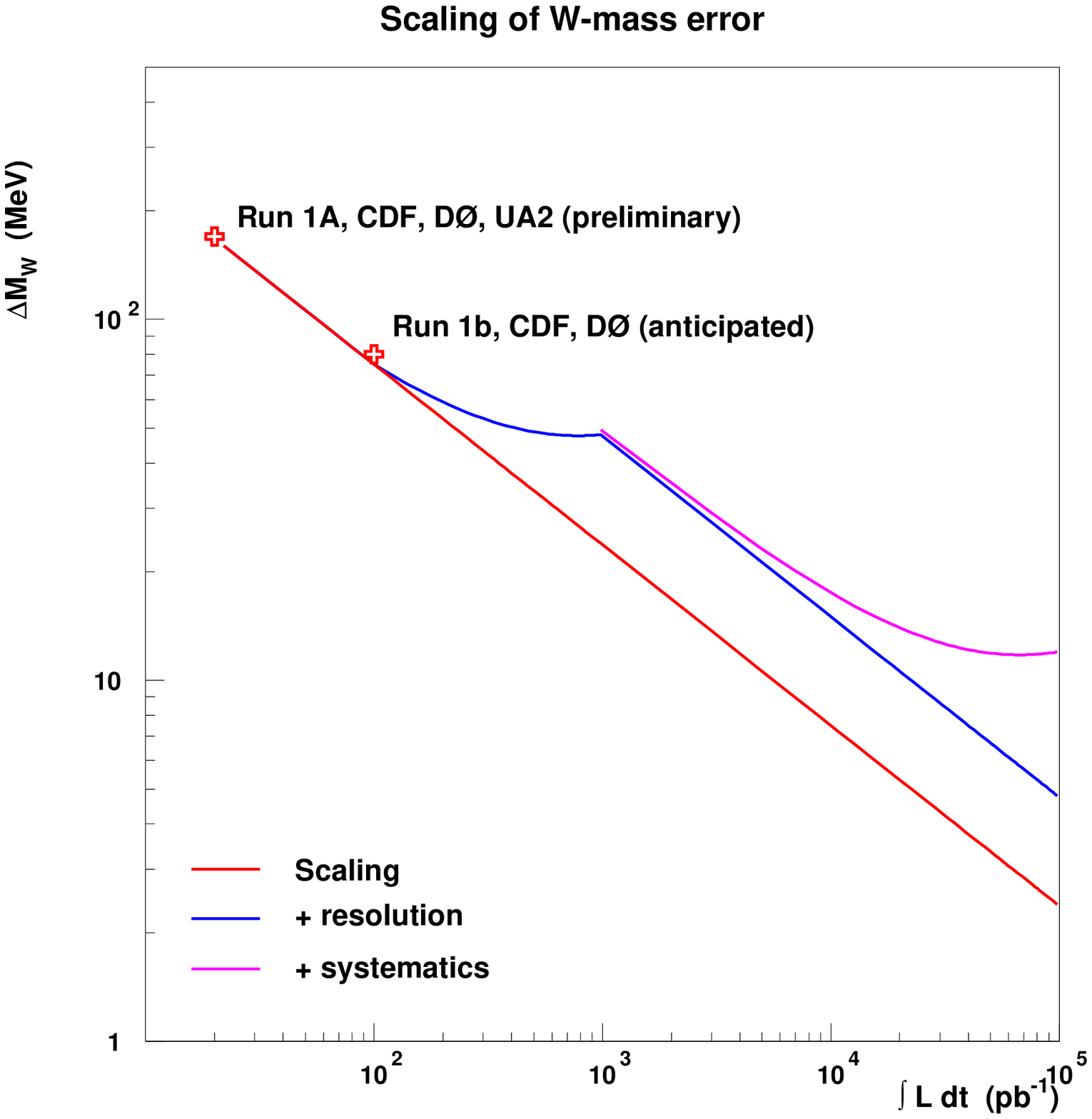, width=10cm}
\end{center}
\caption{Uncertainty on the W mass  measurement as a function of integrated 
luminosity.}
\label{future_mw}
\vspace*{0.3cm}
\begin{center}
\epsfig{file=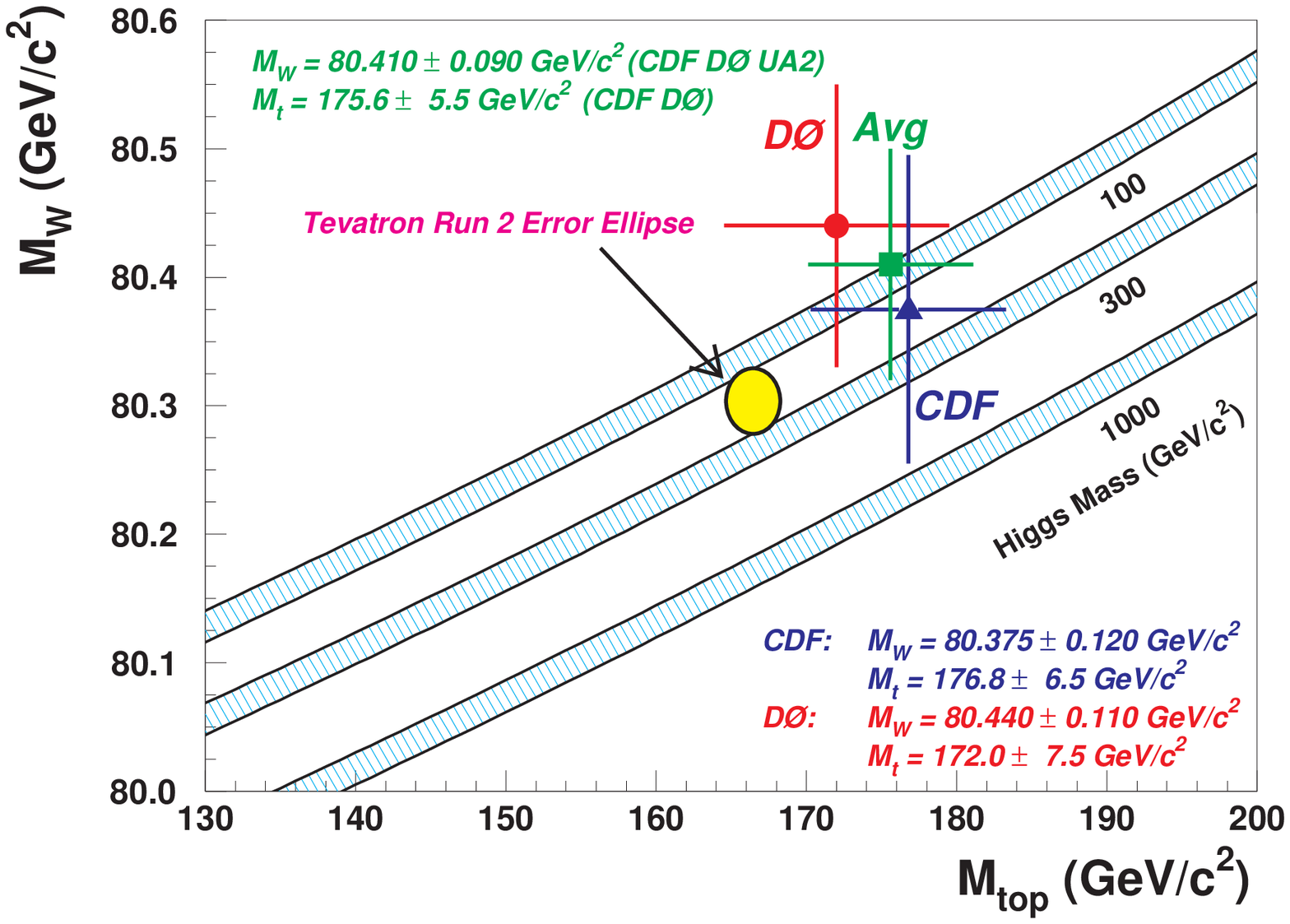,width=13cm}
\end{center}
\vspace*{-0.5cm}
\caption{ The measured  W mass versus top mass in  Run I. 
Theoretical predictions are shown for 
several values of  Standard Model Higgs mass. The  Run II error ellipse is
shown at an arbitrary position.}
\label{mt_mw}
\end{figure}

    More than 1 500 000  events from  $W  \rightarrow e \nu $ decays and about
the same  number from $W  \rightarrow  \mu \nu $  will be  available for
analysis  in  each  experiment.  This  large sample  will  permit the  following
measurements:

 \begin{itemize}

  \item  \textbf{$\delta(M_{W})\simeq 40$ \Mm} (figure~\ref{future_mw}). This  uncertainty combined with
the LEP  measurement and  the  uncertainty on the  top mass  measurement at the
Tevatron will  determine the mass  of the Standard  Model Higgs to a
precision   of       $\delta\textit{M}_{H}/\textit{M}_{H}   \simeq  40\%$ (figure~\ref{mt_mw})

  \item \textbf{W  and Z charge  asymmetry}: constraint  on the choice of the
set  of  parton  density   functions;   measurement  of     $\delta({\sin {^{2}
\theta_{W}}}) \simeq 0.001$.

  \item \textbf{Trilinear gauge bosons  coupling}: Improved limits on anomalous
couplings by a  factor $\simeq 5-15  $ for  WW$\gamma$ and by $\simeq
10-100 $ for ZZ$\gamma$ compared to Run I.

\end {itemize}

\subsection {Search for new phenomena}
  \begin{itemize}
    \item  \textbf{SUSY}\\
Run II will allow the exploration of a large fraction of the MSSM parameter phase space for charginos
($\chi^{\pm}$),    neutralinos  ($\chi^0$), squarks   ($\tilde{q}$) and gluinos
($\tilde{g}$).
      The most promising signals include $\chi^{\pm}$ in trilepton final state,
      $\tilde{g}$ in missing \Et + jets. The mass reach is:
            M$_{\chi^{\pm}}\ \simeq  220$ \Gm; 
             M$_{\tilde{g}} \simeq  400$ \Gm
    \item \textbf{Exotic phenomena}\\
        Search for W'Z', leptoquarks, technicolor, charged Higgs from top decay,
        compositness,..with mass limits improved by a factor 1.5-2 compared to
        RUN I.
  \end {itemize}

\subsection {Properties of the b-quark}
  CDF has already demonstrated in Run I that precision B physics (spectroscopy,
  decays, lifetime, mixing, etc) can be done successfully at the Tevatron.

 The  copious  production of the  various  species of b  hadrons   will allow a
variety  of B  physics  topics  to be    studied~\cite{beaute}  (QCD  tests, 
study of the B$_{c}$ system,
observation of rare decay modes, B$_{s}$  mixing, CP violation). The main focus
will    certainly be  the   search for  CP   violation  and    establishing CKM
constraints: sin$(2\beta)$ will be measured with a precision better
than $\pm 0.14$; from fully  reconstructed B$_{s}$  decays, a value up to x$_{s}
\simeq 20$ could be reached.

%
%
%
%
%
%
\subsection {Properties of QCD }
Precision studies will be made with new probes in new regions of phase space.
Measurements of parton distributions, Drell-Yan
production of W and Z accompanied by jets and non-pertubative phenomena such as
rapidity gaps and diffractive scattering will be of continued importance with 
large sample of data.

\section{Upgrade of the accelerators}

 To  meet the  required  increase  in  luminosity, an upgrade of the  Fermilab
accelerator  complex is underway~\cite{tev}. In  Run II, the
plan is to  deliver an  instantaneous  luminosity  of $2\times 10^{32}$ \Ll. 
For the TeV33 project, an increase  of a factor 5  in instantaneous
luminosity is foreseen.

The key to increase the luminosity is to increase the number of antiprotons. 
This goal is achieved by
replacing  the   existing Main  Ring  with a  new   accelerator, the  {\bf Main
Injector}, and adding a new antiproton storage ring, the {\bf Recycler}, within
a new common  tunnel adjacent to that  of the Tevatron.  This will result in an
overall   increase  both in  bunch  population  and in  the  number of  bunches
available for \pp~collisions.

\begin{itemize} 
\item \textbf{The Main Injector}   
 is  a large aperture, rapid cycling, 120 GeV proton 
synchrotron designed specifically to replace 
the present Main Ring in its two main functions: production of 
$\bar{p}$~'s and injection into the Tevatron.
 In addition  the Main  Injector  will provide an  extracted 120 GeV
beam containing $3\times10^{13}$ protons with a 2.9 second cycle time.

\item \textbf{The Recycler} 
is a  permanent-magnet,  8 GeV  storage ring  which will 
provide a factor of $\simeq$
2 increase in luminosity beyond that projected  with the Main Injector alone.
 \end{itemize}

 Additional changes include:
\begin{itemize}
\item  an increase in beam energy from 900 GeV to the nominal 1 TeV;
\item an increase in the number of bunches from 6 to 36, together with a 
decrease in the crossing time 
from 3400 to 396 ns and later to 132 ns (with a 108 bunches against 108
bunches operation).

\end{itemize}
 The main Tevatron parameters for Run II are compared to the ones in
Run I in table~\ref{tab:run_cond}.
\begin{table}[tbhp]
\begin{center}
\begin{tabular}{|l|l|l|l|}\hline  
                  &  \ \   Run IB                &  \ \   Run II                
  &\qquad units \\ \hline
Protons/bunch   &    2.32$\times 10^{11}$  & 2.70$\times 10^{11}$ &\\
Antiprotons/bunch&    5.50$\times 10^{10}$  &  7.50$\times 10^{10}$ & \\
Total Antiprotons &    3.30$\times 10^{11}$   &      1.98$\times 10^{12}$ &\\
Energy/beam             &  900        & 1000       &     GeV \\ 
Number of bunches &    6 + 6         &           36 + 36 &\\
Antiprotons stacking  &    0.6$\times 10^{11}$&2$\times 10^{11}$ &     per hour \\
Typical Luminosity &    1.6$\times 10^{31}$  &       2.0$\times 10^{32}$ &cm
$^{-2}$s$^{-1}$ \\
Integrated Luminosity &    3.2   &      41.0       &      pb$^{-1}$per week  \\
Bunch spacing &     3500       &       396 $\ra$ 132      &        ns \\
Interactions per crossing &  2.7   &      5.8 $\ra$ 2.0&     \\  \hline
\end{tabular}
\caption{ Tevatron running conditions for Run II compared to Run I }
\label{tab:run_cond}
\end{center}
\end{table}

\section{Detector Upgrade}

The upgrade builds on existing strength to identify and measure leptons, photons and jets, with a nearly complete solid angle coverage for calorimetry and muons detection.

The high luminosity and bunch spacing as well as the increase in detector occupancy require extensive changes in the detector of Run I.
Higher level of radiation requires extensive shielding too. Computing and data storage systems must be able to handle a $\simeq$ 40  fold increase in event collection with respect to Run I.

An overall view of the \Dzero detector is shown on figure~\ref{d0_det}, with
primary detector systems still remaining: the liquid argon calorimeter and the central part of the muon detector system. The major element of the upgrade is the replacement of the non-magnetic inner tracking system with a high precision integrated tracker including a solenoid~\cite{det}.

\subsection{Central Tracking System}

The new high precision compact tracker enhances the identification of electrons and muons, allows the in situ calibration of the calorimetric energy scale and makes the whole area of B physics studies accessible to \Dzero.

\begin{figure}
\begin{center}
\epsfig{file=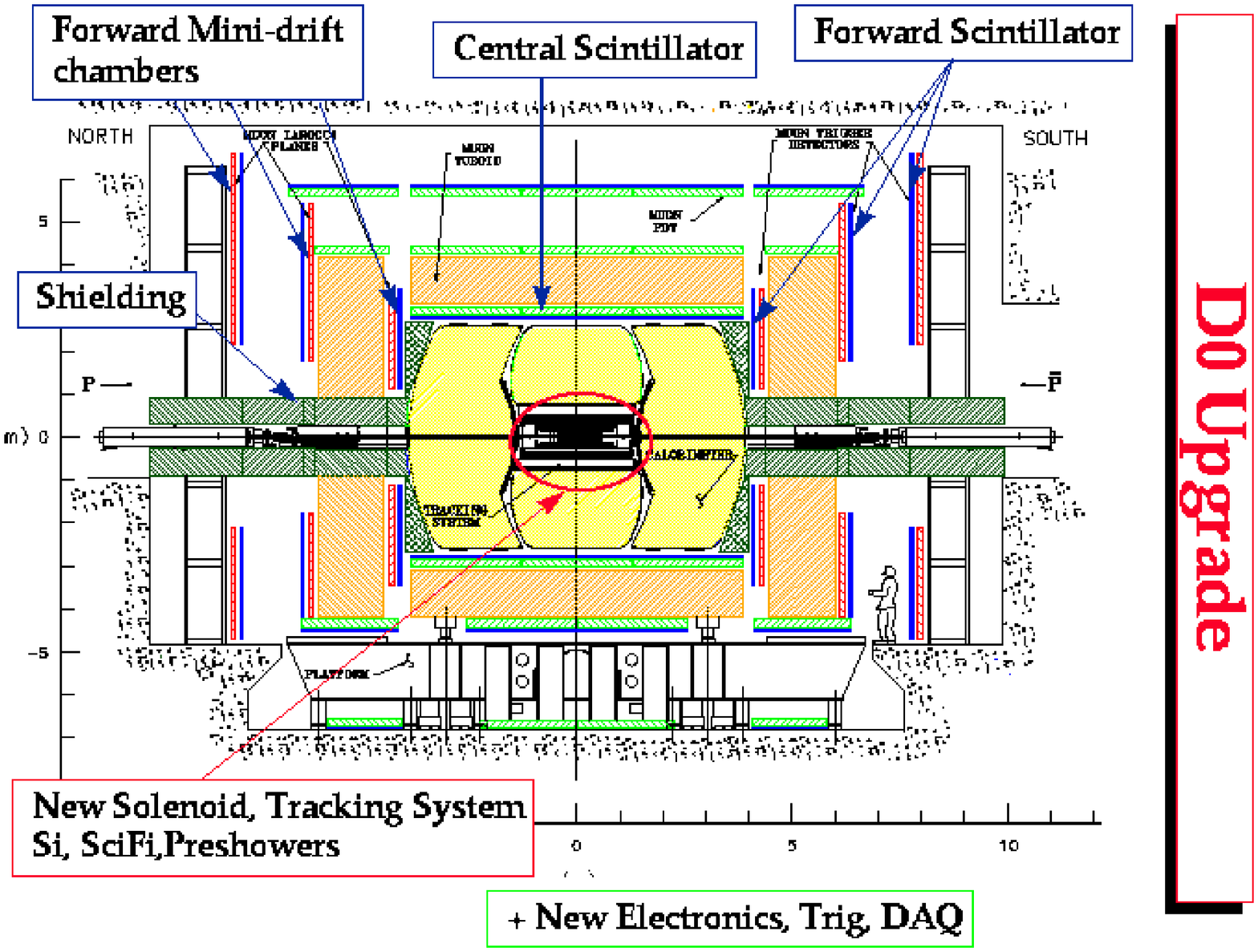,angle=-90,bb=0 0 610 842,width=14cm}
\end{center}
\caption{ The \Dzero detector.}
\label{d0_det}
\vspace*{1cm}
\begin{center}
\epsfig{file=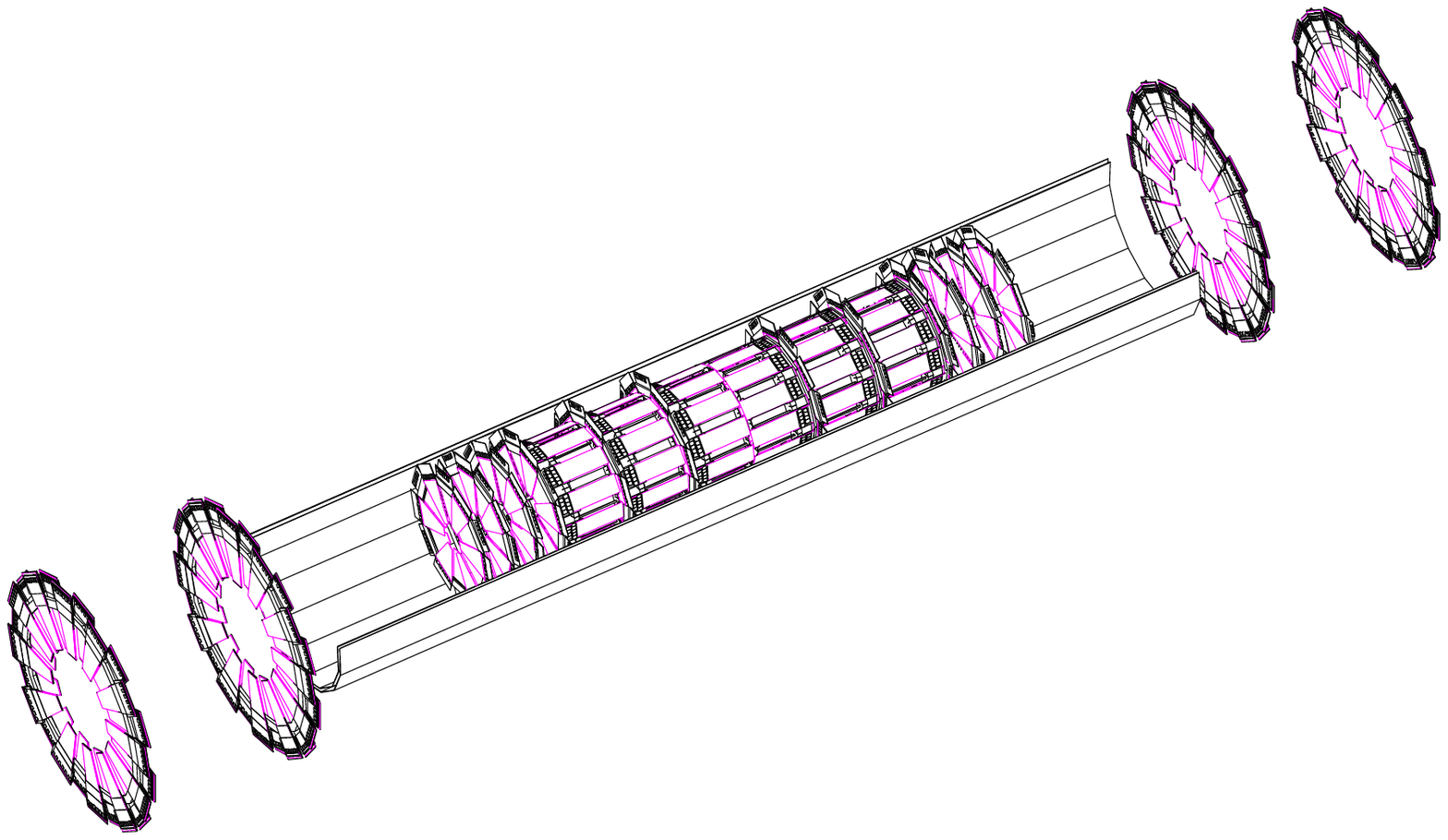,width=12cm}
\end{center}
\caption{ The Silicon Microvertex Tracker.}
\label{d0_track}
\end{figure}

The central tracking detector consists of:

\begin{itemize}

\item A Silicon Microstrip Tracker (\textbf{SMT}): 6 barrels of silicon detectors with 3D readout interspersed with 14 z-disks that extend out to \etae=3 (figure~\ref{d0_track}). Staggered planes in the SVX barrels overlap such as to yield an average of 5.5 layers on track. 
The single point resolution has been shown to be $\leq$ 10 $\mu$m.

Electronic readout will be done using the 128 channel SVX II readout chip developed by Fermilab and Lawrence Berkeley Laboratory. Each channel contains a charge sensitive preamp, 32 stages of analog pipeline delay, an 8 bit analog to digital converter, and sparse data readout.

The \textbf{SMT} is equipped with 840,000 channels.
 
 \item A Central Fiber Tracker (\textbf{CFT}): eight superlayers of scintillating fibers, each arranged as two layers axial and stereo of staggered fiber doublets. The fibers are mated to 8-11 m multiclad clear fiber waveguides which conduct the light to photodetectors. 

The photodetectors are capable of detecting single photons with high efficiency at high rates and with large gain. For the first time in a high energy physics experiment, a large number (77,000) of visible light photon counters (VLPC's) are used. VLPC's are impurity band conduction devices derived from solid state photomultipliers. Test results indicate they can detect single photons with quantum efficiencies up to 80\% and gains $\ge$ 10$^4$. A cosmic ray test has confirmed that single point resolution for doublet layers are $\simeq$ 100 $\mu$m with hit efficiencies of 99.5 \%.

\item A superconducting solenoid, 2.8 meter long, 50 cm in diameter, installed inside the central calorimeter cavity. 

\item Preshower detectors made of scintillating strips equipped with wavelength shifter readout. The central preshower is mounted between the magnet coil and the outer wall of the central cryostat while the forward preshowers are mounted on the front face of the forward cryostats.

\end{itemize}

\subsection{Calorimeter} 

The Uranium-Liquid argon calorimeter is intrinsically radiation hard, and no upgrade or modifications are needed, except for the front end electronics which has to be completely changed due to shortened bunch crossing intervals.

\subsection{Muon System} 

One of the strong feature of the \Dzero detector is its almost complete muon coverage. 
In the central region (\etae $\leq$ 1), three superlayers of proportional drift tubes (\textbf{PDT}), one inside and two outside of a 1.5 m shield of a toroidally magnetized iron, allow to measure the muon momentum.  Two layers of scintillators are used in the trigger. 
In the forward direction, \textbf{PDT} are replaced with mini-drift tubes (\textbf{MDT}). Three layers of highly segmented scintillator counters provide the trigger.
 
Of particular B physics relevance is the lowered muon $p_T$ threshold from 4 GeV/c in Run I to 1.5 GeV/c in Run II. Due to the presence of a central magnetic spectrometer, the momentum resolution is significantly enhanced to Run I capabilities.


\section{Trigger System}

The trigger system has to cope with a 10 fold higher luminosity and a $\sim$
40 fold decrease in the time between beam crossing (from 4 $\mu$s to 132 ns).
These two parameters require significant upgrade of the entire trigger system.

The first trigger stage (L1) includes scintillating, tracking and calorimeter detectors. It provides a trigger decision in less than 4.2 $\mu$s with a rate of 10 kHz.

The second stage (L2) comprises preprocessors to reorganize the detector and L1 trigger information and a global processor to test for correlation between L1 triggers. The L2 trigger has a rate of 1 kHz at a maximum deadtime of 5 \%.

The third level (L3) is a conventional multiprocessor farm, where each node receives one event at a time and performs full reconstruction. A 20 Hz output is written into a massive storage system.

\section {Beyond Run II: TeV33}

 Run II  promises a rich and  varied physics  program at the  Tevatron. However
there  remains  much  interesting high  \pt~physics which  require  integrated
luminosities with an  order of magnitude  larger, $\simeq$  30 \Fb. As an example, it
may be possible to observe light and  intermediate Higgs bosons in a mass range
\mbox{$80\leq$  M$_{\textit{H}} \leq  130$ \Gm}. Of  particular  interest would be the
measurement of the W mass to \mbox{20 \Mm} and of the top mass to \mbox{1 \Gm}.

 The challenge is to trigger and record data with an instantaneous luminosity of
\mbox{$1\times  10^{33}$~\Ll}.  Preliminary  studies have  shown that  this is feasible
providing upgrade  detectors beyond the  Run II upgrades.  The best strategy for
running during the TeV33 era is under discussion.

\section{Conclusion}

The \Dzero detector upgrade consisting of a new tracking system, an improved muon detection system, new triggers, and upgraded electronics, is on schedule to be completed for data taking in spring 2000.

The energy frontier will remain at the Tevatron until the middle of the next decade. The higher integrated luminosity and the increase in sensitivity of the upgraded \Dzero detector will allow the full exploitation of a rich program of physics. Many precision measurements will provide stringent tests of the Standard Model and possibly give some hints of deviation from its predictions. There exist good chances to either discover new physics or, in its absence, to severely constrain the extension to the Standard Model.

\section*{Acknowledgments}

 I would like to thank the organizers for the invitation to this conference
and my colleagues from LAL for helping me in the preparation of this talk.


\begin{thebibliography}{99}
\bibitem{tev2000}
Future Electroweak Physics at the Fermilab Tevatron Report of the tev-2000 study group, FERMILAB-Pub-96/082.
\bibitem{beaute} Prospects for Mixing and CP Violation at the Tevatron, 
FERMILAB-conf-96/274-E.
\bibitem{tev}
Tevatron Status and Future Plans FERMILAB Conf-96-408.
\bibitem{det}
``The \Dzero Upgrade: The detector and its Physics'' \Dzero collaboration, FERMILAB Pub-96/357-E (1996).



\end{thebibliography}
\end{document}